\documentclass[prl,twocolumn,superscriptaddress,showpacs]{revtex4}

\usepackage{graphicx}
\usepackage{dcolumn}
\usepackage{amsmath,tabularx}

\newcommand\UBasel{Universit\"{a}t Basel, CH-4056 Basel, Switzerland}
\newcommand\UVa{University of Virginia, Charlottesville 22903}
\newcommand\Yerevan{Yerevan Physics Institute, Yerevan, Armenia}
\newcommand\NCAT{North Carolina A\&T State University, Greensboro,
  North Carolina 27411}
\newcommand\TelAviv{Tel Aviv University, Tel Aviv, 69978 Israel}
\newcommand\FIU{Florida International University, Miami, Florida 33199}
\newcommand\UMD{University of Maryland, College Park, Maryland 20742}
\newcommand\OhioU{Ohio University, Athens, Ohio 45071}
\newcommand\Jlab{Thomas Jefferson National Accelerator Facility,
  Newport News, Virginia 23606 }
\newcommand\MissSU{Mississippi State University, Mississippi State,
  Mississipi 39762}
\newcommand\HamptonU{Hampton University, Hampton, Virginia 23668}
\newcommand\SUNO{Southern University at New Orleans, New Orleans,
  Louisiana 70126}
\newcommand\LaTech{Louisiana Tech University, Ruston, Louisiana 71272}
\newcommand\NCCU{North Carolina Central University, Durham, North
  Carolina 27707}
\newcommand\NSU{Norfolk State University, Norfolk, Virginia 23504}
\newcommand\ODU{Old Dominion University, Norfolk, Virginia 23529}
\newcommand\VaTech{Virginia Polytechnic Institute \& State University,
Blacksburg, Virginia 24061} \newcommand\Duke{Duke University and
TUNL, Durham, North Carolina
27708} \newcommand\NIKHEF{Vrije Universiteit, 1081 HV Amsterdam, The
Netherlands}

\newcommand\GeVcsqr{\mbox{(GeV/c)$^2$}}
\newcommand\GEn{\mbox{$G_{\!E}^{\,n}$}}
\newcommand\GMn{\mbox{$G_{\!M}^{\,n}$}}

\begin{document}

\preprint{HEP/123-qed}

\title[Short Title]{Measurement of the Electric Form Factor of the
  Neutron at $Q^2=0.5$ and 1.0
  GeV$^2$/c$^2$}

\author{G.~Warren}	\affiliation{\UBasel}\affiliation{\Jlab}
\author{F.~Wesselmann}	\affiliation{\UVa}
\author{H.~Zhu}		\affiliation{\UVa}
\author{P.~McKee}	\affiliation{\UVa}
\author{N.~Savvinov}	\affiliation{\UMD}
\author{M.~Zeier}	\affiliation{\UVa}
\author{A.~Aghalaryan}	\affiliation{\Yerevan}
\author{A.~Ahmidouch}	\affiliation{\NCAT}
\author{H.~Arenh\"{o}vel}\affiliation{Johannes
Gutenberg-Universit\"{a}t, D-55099 Mainz, Germany}
\author{R.~Asaturyan}	\affiliation{\Yerevan}
% ...

\collaboration{The Jefferson Lab E93-026 Collaboration}
\date{January 30, 2004}

\begin{abstract}
The electric form factor of the neutron was determined from
measurements of the $\vec{d}\left(\vec{e},e^\prime n\right)p$ reaction
for quasielastic kinematics. Polarized electrons were scattered off a
polarized deuterated ammonia ($^{15}$ND$_3$) target in which the
deuteron polarization was perpendicular to the momentum transfer.  The
scattered electrons were detected in a magnetic spectrometer in
coincidence with neutrons in a large solid angle detector.  We find
$\GEn{} = 0.0526 \pm 0.0033 (stat) \pm 0.0026 (sys)$ and $0.0454 \pm
0.0054 \pm 0.0037$ at $Q^2 = 0.5$ and 1.0~\GeVcsqr, respectively.
\end{abstract}


\maketitle


The electric form factor of the neutron \GEn{} is a fundamental
quantity in nuclear physics.  Knowledge of \GEn{} over a large range
of momentum transfer leads to insight to physics beyond the simple
SU(6) symmetric models, for which it must vanish.  Beyond nucleon
structure, our understanding of \GEn{} has an impact on a broad range
of topics, which vary from the charge radii of
nuclei~\cite{Friar:pp} to extracting the strangeness content of the
nucleon~\cite{Kumar:eq}.

Historically, measurements of \GEn{} have suffered from large
uncertainties due to experimental technique and nuclear model
dependence.  Early attempts to measure \GEn{} from cross sections
sometimes failed to determine even the sign.  Until recently, the
best determination of \GEn{} came from elastic electron-deuteron
measurements, but the errors were large, $\sim$30\%, due to their
dependence on the nucleon-nucleon potential model~\cite{platchkov90}.
The first polarization measurements of \GEn{}, conducted with $^2$H and
$^3$He targets, differed significantly because final state interactions
were not addressed~\cite{becker99,ostrik99}.  
These experiences highlight the
importance of measuring \GEn{} using different reactions.  

In the last few years the experimental understanding of \GEn{} has
improved considerably.
The results from polarization experiments described in
Ref.~\cite{HG01}-\nocite{passchier99,golak01,eden94,herberg99}\cite{bermuth03}
are consistent and provide good accuracy in the kinematic region of
the four-momentum transferred squared $Q^2\le 0.7$~\GeVcsqr{}
(henceforth units of $Q^2$ are assumed to be \GeVcsqr).  Until now,
values of \GEn{} determined from the deuteron quadrupole form factor
\cite{SS02} provide the only information in the kinematic region above
$Q^2\ge 0.7$ and leave the $Q^2>1.6$ region undetermined.

This letter describes a first measurement of \GEn{} at
$Q^2=1.0$ using a polarized target. In addition, the result
for $Q^2=0.5$ has half the uncertainty as the previous
measurement~\cite{HG01} and is the most precise result near the peak
of $\GEn{}$.  These results are largely insensitive to the model of
the nucleon-nucleon (NN) potential so that, compared to those of
Ref.~\cite{SS02}, they are more reliable.

To determine \GEn{}, the helicity dependent rate asymmetry in electron
scattering was measured.  In the ideal case of a polarized electron
scattering elastically off a free polarized neutron, with the neutron
polarization vector in the scattering plane and perpendicular to the
momentum transfer $\vec q$, \GEn{} is related to the beam-target
asymmetry term $A^V_{en}$~\cite{donnelly} by
\begin{equation}A^V_{en} = 
\frac{-2\sqrt{\tau(\tau+1)} \tan(\theta_e/2)\GEn\GMn}{(\GEn)^2 +
\tau[1+2(1+\tau)\tan^2(\theta_e/2)](\GMn)^2 },
\end{equation} 
where $\tau = Q^2/4M^2_n$, $M_n$ is the mass of the neutron,
\GMn{} is the magnetic form factor of the neutron, and
$\theta_e$ is the electron scattering angle. 

For lack of a free neutron target, the actual measurements were
performed on a polarized deuterium target.  The measured experimental
asymmetries were due to a combination of several physics asymmetries
scaled by the electron and target vector and tensor polarizations (that 
are described in detail in our previous work \cite{HG01}, not repeated 
here for brevity's sake).
The proper averaging of the asymmetry (symmetrically around $\vec{q}$)
and the negligible contributions from the target tensor asymmetry
simplify the relationship of the measured asymmetry $\epsilon$ to the
deuteron vector asymmetry $A^V_{ed}$ so that $\epsilon = f P_e P_1^d
A^V_{ed}$, where $P_e$ is the beam polarization, $P_1^d$ is the
deuteron vector polarization and $f$ is the dilution factor due to
scattering from nucleons other than polarized deuterons in the target.
Calculations show that $A^V_{ed}$ depends linearly on \GEn{} for the
kinematics of the experiment~\cite{arenhoevel88}.

The measurements were conducted in Hall C of the Thomas Jefferson
National Accelerator Facility in a setup similar to that of the
previous measurement~\cite{HG01}. The longitudinally polarized
electron beam~\cite{source} was scattered off a polarized frozen
deuterated ammonia ($^{15}$ND$_3$) target.  The scattered electrons
were detected by the High Momentum Spectrometer (HMS), and the
neutrons were detected by a dedicated neutron detector.  The central
kinematics for the two measurements as well as the average beam and
target polarizations are listed in Table~\ref{tab:kinematics}.
The average deuteron luminosity was $10^{35}\!/\!cm^{2}\!/\!s$.

The polarized target~\cite{target} consisted of ammonia granules
submerged in liquid He and maintained at 1 K by a $^4$He evaporation
refrigerator. The target spins were aligned by a 5 T magnetic field
generated by a pair of superconducting coils.  The polarization was
enhanced via dynamic nuclear polarization~\cite{Crabb94} and measured
with a continuous-wave NMR system~\cite{nmr}.  To minimize localized
heating and depolarization of the target material, the electron beam
was rastered uniformly in a circular pattern with a 1~cm radius.

A two-magnet chicane compensated for the deflection of the electron
beam by the target field.  After traversing the target, the beam
passed through a helium bag to a special dump in the experimental
hall.  The polarization of the beam was measured at regular intervals
throughout the experiment with a M\o{}ller polarimeter~\cite{moller}.
The beam helicity was changed in a pseudorandom sequence 30 times per
second to minimize charge fluctuations correlated with helicity.

The HMS was operated in its standard mode for the detection of
electrons.  The established reconstruction algorithms were augmented
to account for the large beam rastering and the effects of the target
magnetic field on the scattered electrons.

\begin{table}
\begin{center}
\begin{tabularx}{0.43\textwidth}{cccccccc}
\hline
\hline
\rule{0cm}{9 pt} 
$Q^2$ & $E$ & $E^\prime$ & $\theta_e$ &  $\theta_n$ & $\theta_B$ & 
$\left<P_e\right>$ & $\left<P_t\right>$ \\ 
\GeVcsqr{} & (GeV) & (GeV) & (deg) &(deg)& (deg)& (\%)& (\%) \\
\hline 
0.5 &2.331 & 1.963 & -18.5 & 60.5 & 150.4 & 78.2 & 24.1 \\ 
1.0& 3.481 & 2.810 & -18.0 & 53.3 & 143.3 & 71.8 & 23.8 \\ 
\hline
\hline
\end{tabularx}\end{center}
\caption{Central kinematics and average polarizations. $E$
  ($E^\prime$) is the energy of the incident (scattered) electron.
  $\theta_e$ ($\theta_n$) is the angle of the scattered electron
  (neutron).  $\theta_B$ is the orientation of the target polarization
  axis.  $\left<P_e\right>$ and $\left<P_t\right>$ are the average
  beam and target polarizations, respectively}\label{tab:kinematics}
\end{table}

The neutron detector consisted of multiple vertical planes of
segmented plastic scintillators.  Two planes of thin scintillators
served to distinguish charged particles.  Behind these were six planes
of thick scintillators to detect the neutrons.  All scintillators were
equipped with photomultipliers on both ends to provide spatial and
time information for the detected particle.  The 88 thick
scintillators provided a neutron detection volume that was roughly 160
cm wide, 160 cm tall and 90 cm deep.  The front of the detection
volume was approximately 4.2~m and 6.2~m from the target during the
$Q^2 = 0.5$ and 1.0 measurements, respectively. The detector was
shielded from direct gamma rays from the target by a 2.5 cm lead
curtain, and the entire assembly was housed in a thick-walled concrete
hut, which was open to the target.

The trigger was set up so that the neutron detector was read out for
every electron trigger in the HMS.  Coincidences between the electrons
and the knock-out nucleon were determined offline. 

The experiment was simulated using Monte Carlo (MC) techniques.  The
simulation package, based on MCEEP~\cite{MCEEP}, included the charged
particle transport through the target's magnetic field and the optical
and aperture model of the HMS from the Hall C simulation package
SIMC~\cite{SIMC}.  The MC served two principal functions: to determine
the dilution factor and to average the theoretical asymmetries over the
acceptance.  For the dilution factor, all target materials were
included in the simulation: the deuterium and the nitrogen in the
ammonia, the liquid helium bath, the aluminum target cell windows and
the NMR coil.

Contributions from electron-neutron events originating from
quasielastic scattering and pion production were included in the MC
for all target materials.  It was found that the $(e,e^\prime
n)\pi$ contribution to the event sample was negligible for
$Q^2=0.5$ and less than 0.5\% for $Q^2=1.0$.  Two-body knockout
contributions were examined in the MC (following 
\cite{ryckebusch}) and were also negligible.

\begin{figure}
\vskip -0.5 in
\begin{center}
\includegraphics[height=3.8 in]{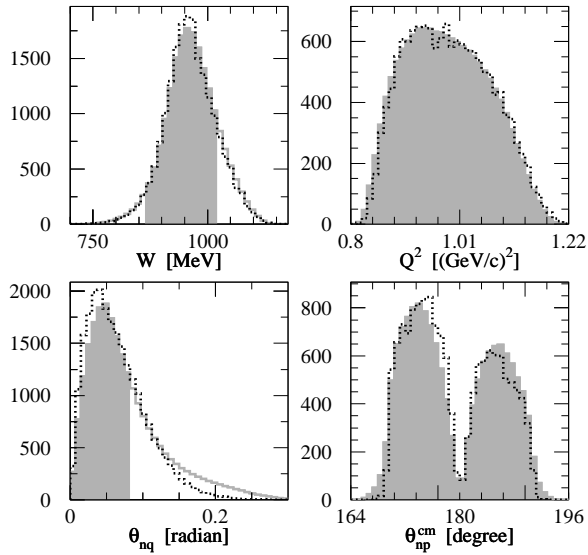}
\vskip -0.2 in
\caption{Comparison of $Q^2=1.0$ MC and data electron-neutron event 
  distributions from all materials in the target for
  four kinematic variables: invariant mass $W$, $Q^2$, angle between
  the neutron and $\vec{q}$ in lab frame $\theta_{nq}$, and angle
  between neutron-proton system and the momentum transfer in the
  center-of-momentum frame $\theta_{np}^{cm}$.  The solid grey
  histograms correspond to the data, and the dotted black histograms
  correspond to the simulation.  Only the shaded regions, dominated by 
  events from $^2$H were used to determine \GEn{}.
  \label{fig:mc-data}}
\end{center}
\vskip -0.2 in
\end{figure}

The simulated rates were normalized to the measured ammonia rates,
and the variations in the ratios of MC rate to observed rate for
various target materials were used to determine the uncertainty in the
dilution factor.  A comparison of event distributions of the data
and MC for several kinematic variables for $Q^2=1.0$ is shown in
Fig.~\ref{fig:mc-data}; agreement is very good.  The comparison for
the $Q^2=0.5$ data is similar and slightly better than in the previous
experiment~\cite{HG01}.

Several corrections to the measured asymmetry were applied.  The
leading correction was for charge exchange of protons in the lead
shielding: $-3.8 \pm 1.1\%$ ($-3.3 \pm 3.0\%$) for
$Q^2=0.5$ ($Q^2=1.0$).  The charge exchange in the target
material itself was small: $-0.3 \pm 0.3\%$ ($-0.3 \pm 0.3\%$).  The
accidental background rate was $1.9 \pm 0.1\%$ ($0.5 \pm 0.2\%$) with
no statistically significant asymmetry.  Internal radiation effects on
the asymmetry were also corrected: $0.9 \pm 0.5\%$ ($0.6 \pm 0.5\%$).
The effect of external radiation was insignificant compared to the 0.5\%
MC statistical uncertainty.  The contamination from neutral
pions generated by protons in the lead shielding of the neutron
detector was also found to be insignificant.

The physics model of $A^V_{ed}$ used in the MC was based on the
calculations of Arenh\"ovel, Leidemann and
Tomusiak~\cite{arenhoevel88}.  It included a non-relativistic
description of the $n-p$ system in the deuteron using the Bonn
$R$-space $NN$ potential~\cite{bonn} for both the bound state and the
final state interactions.  The full calculation included meson
exchange, isobar configuration currents and relativistic corrections.

The model assumed a scaled Galster parameterization~\cite{galster} for
\GEn{} and the dipole parameterization for \GMn.  The Galster
parameterization is 
$ 
\GEn{}(Q^2) = -\mu_{n}\tau/(1+5.6\tau)G_D(Q^2), 
$
where $G_D = 1/(1+Q^2/0.71)^2$ is the dipole form factor.  Various
scale factors of this parameterization, ranging from 0.5 to 1.5,
were examined.  The potential impact of the $Q^2$ dependence of the
\GEn{} parameterization was found to be negligible because the $Q^2$
acceptance was not large and the $Q^2$ dependence of \GEn{} over the
acceptance was mostly linear.  The acceptance-averaged value of
$A^V_{ed}$ has a linear dependence on \GMn, so one can trivially
correct for more accurate \GMn{} values.

The value of \GEn{} was determined by comparing the acceptance
averaged $A^V_{ed}$ of the data to that of the MC.  The theoretical
asymmetries were determined for a range of scaling factors of the
Galster parameterization to determine the corresponding \GEn:
$\GEn{}/\mathrm{Galster} =1.003 \pm 0.064$ and $1.172 \pm 0.140$ for
$Q^2=0.5$ and 1.0, respectively.  The uncertainties are statistical
only.  To account for the slight deviations of \GMn{} from the dipole
form factor, the recent fit to the \GMn{} data~\cite{baselgmn} was
used: $\GMn/\mu_{n} G_D = 1.007 \pm 0.005$ and
$1.072 \pm 0.014$ for $Q^2=0.5$ and 1.0, respectively.  Then the
values for \GEn{} are:
\begin{eqnarray*}
  \GEn{}(Q^2 = 0.5) & = & 0.0526 \pm 0.0033 \pm 0.0026, \\
  \GEn{}(Q^2 = 1.0) & = & 0.0454 \pm 0.0054 \pm 0.0037, 
\end{eqnarray*}
where the first uncertainty is statistical and the second is
systematic.  The $Q^2 = 0.5$ result agrees well with the previous
result reported in Ref.~\cite{HG01}.

Other systematic uncertainties for $Q^2=0.5$ ($Q^2=1.0$) were:
dilution factor, 3.4\% (3.0\%); target polarization, 2.9\% (4.6\%);
central kinematic values, 1.2\% (3.4\%); beam polarization, 1.1\%
(3.3\%).

\begin{figure}
\begin{center}
\includegraphics[width=3.9 in]{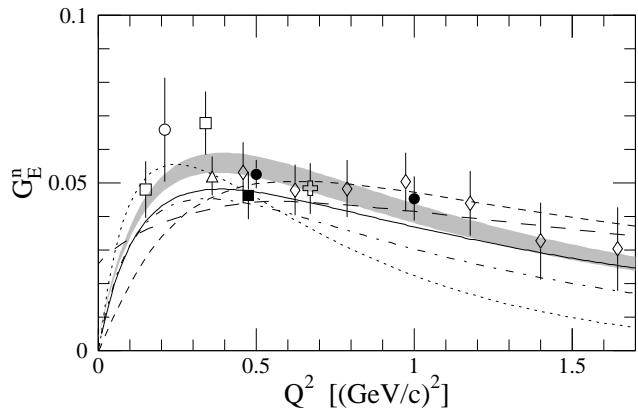}
\vskip -0.2 in
\caption{Comparison of this experiment with data from recent
measurements.  The data points are: diamonds~\cite{SS02}, hollow
squares~\cite{herberg99}, triangle~\cite{becker99}, hollow
circle~\cite{passchier99}, solid square~\cite{HG01}, plus
sign~\cite{bermuth03}.  Solid circles: this paper.  See text for a
description of the curves.  \label{fig:genresults}}
\end{center}
\vskip -0.3 in
\end{figure}

The results as compared to recent measurements are shown in
Fig.~\ref{fig:genresults}.  The new data at $Q^2= 1.0$ provides the
important experimental confirmation of the decline of \GEn{} following
the Galster form.  The data shown in the figure were fit to the form
$ 
\GEn{}(Q^2) = -\mu_{n} a\tau/(1+p\tau)G_D(Q^2), 
$ 
The parameter $a = 0.895
\pm 0.039$, the slope at $Q^2=0$, is fixed by atomic
measurements of the neutron charge radius~\cite{kopecky}.  The fit yielded
$p = 3.69 \pm 0.40$.  The errors obtained
for the fit parameters are uncorrelated.  The one-sigma error region
of the fit is shown in Fig.~\ref{fig:genresults} as the shaded band.

Many recent models ~\cite{lomon}--\nocite{lightfront,holzwarth,pfsa}\cite{miller} 
have attempted to predict or fit the nucleon electromagnetic form factors. 
Fig.~\ref{fig:genresults} compares the data with  two recent calculations that use 
covariant formulations of the constituent quark model with quark-quark interactions 
fitted to spectroscopic data.  The point-form spectator approximation (PFSA) of 
\cite{pfsa} (dot-dash) uses a Goldstone boson exchange interaction with pointlike 
constituent quarks while the light-front (LF) calculation of \cite{lightfront} 
(short dash) uses a one-gluon exchange interaction with constituent quark form factors 
fitted to data for $Q^2 < 1$.  The use of constituent form factors improves the fit to 
the nucleon magnetic form factors at larger $Q^2$, but the PFSA seems to describe \GEn{} 
better at low $Q^2$ with fewer parameters.

    Also shown in Fig.~\ref{fig:genresults} are the results from a hybrid
model that interpolates between vector-meson dominance at low $Q^2$
and perturbative QCD at high $Q^2$~\cite{lomon}(solid line), from a light front model 
where the nucleon is considered a system of three bound quarks surrounded by a cloud of 
pions\cite{miller} (long dash) and   from a soliton model\cite{holzwarth} (dotted) 
whose basic features include an extended object, partial coupling to the vector mesons 
and relativistic recoil corrections.
  While all these models agree qualitatively
with the data, none agree with the data for the entire range of $Q^2$.

It is remarkable that in  the last five years, the experimental precision in \GEn{} 
measurements has improved to the 10\% level.
 This significant improvement provides a rigorous
challenge for models of the nucleon structure because this
electromagnetic form factor is the most sensitive to physics beyond
the simplistic SU(6) symmetric picture.

In conclusion, the electric form factor of the neutron at $Q^2=0.5$
and $1.0$ has been determined from measurements of the beam-target
asymmetry.  This experiment provides the highest $Q^2$ measurement to
date using a polarized target and the most precise measurement near
the maximum of \GEn{}.  

We wish to thank the Hall C technical and engineering staff at TJNAF
as well as the injector, target and survey groups for their
outstanding support.  This work was supported by the Commonwealth of
Virginia, the Schweizerische Nationalfonds, the U.S. Department of
Energy, the U.S. National Science Foundation, the U.S.-Israel
Binational Science Foundation and the Deutsche Forschungsgemeinshaft.
The Southeastern University Research Association (SURA) operates the
Thomas Jefferson National Accelerator Facility for the U.S. Department
of Energy under contract DE-AC05-84ER40150.


\begin{thebibliography}{99}
\bibitem{Friar:pp} J.L. Friar and J.W. Negele,
    Advances in Nuclear Physics, Vol. 8, 219, 1975.
\bibitem{Kumar:eq} K.S. Kumar and P.A. Souder,
    Prog. Part. Nucl. Phys. {\bf 45}, S333 (2000).
\bibitem{platchkov90} S. Platchkov {\it et al.}, 
    Nucl. Phys. {\bf A510}, 740 (1990).
\bibitem{becker99} J. Becker {\it et al.}, 
    Eur. Phys. J. A {\bf 6}, 329 (1999).
\bibitem{ostrik99} M. Ostrick {\it et al.}, 
    Phys. Rev. Lett {\bf 83}, 276 (1999).
\bibitem{HG01} H. Zhu {\it et al.}, Phys. Rev. Lett. {\bf 87}, 081801
  (2001).
\bibitem{passchier99} I. Passchier {\it et al.}, Phys. Rev. Lett. {\bf 82}, 4988 (1999).
\bibitem{golak01} J. Golak {\it et al.}, Phys. Rev. C {\bf 63},
  034006 (2001). 
%
%
\bibitem{eden94} T. Eden {\it et al.}, Phys. Rev. C {\bf 50}, R1749
  (1994).
\bibitem{herberg99} C. Herberg {\it et al.}, Eur. Phys. J. A {\bf 5},
  131 (1999). 
\bibitem{bermuth03} J. Bermuth {\it et al.}, Phys Lett B. {\bf 564},
  199 (2003).  D. Rohe {\it et al.}, Phys. Rev. Lett. {\bf 83}, 4257 (1999).
\bibitem{SS02} R. Schiavilla and I. Sick, 
    Phys. Rev. C {\bf 64}, 041002(R) (2002).
\bibitem{donnelly} T.W. Donnelly and A.S. Raskin, Ann. Phys. (N.Y.)
  {\bf 169} 247 (1986). A.S. Raskin and T.W. Donnelly,
  Ann. Phys. (N.Y.) {\bf 191}, 78 (1989).
\bibitem{arenhoevel88} H. Arenh\"ovel, W. Leidemann, E.L. Tomusiak,
  Z. Phys A {\bf 331}, 123 (1988); {\bf 334}, 363(E) (1989).
  H. Arenh\"ovel, W. Leidemann, E.L. Tomusiak, Phys. Rev. C {\bf 46},
  455 (1992).  H. Arenh\"ovel, private communication.
\bibitem{source} C. Sinclair, TJNAF Report No. TJNAF-TN-97-021.
\bibitem{target} D. Crabb and D. Day, Nucl. Inst. Meth. A {\bf 356}, 9
  (1995).  T. D. Averett {\it et al.} Nucl. Inst. Meth. A {\bf 427},
  440 (1999).
\bibitem{Crabb94} D.G. Crabb and W. Meyer, Ann. Rev. Nucl. Part. Sci. 47, 67 (1997).
\bibitem{nmr} G. Court, Nucl. Inst. Meth A {\bf 324}, 433 (1993).
\bibitem{moller} M. Hauger {\it et al.}, Nucl. Inst. Meth. A {\bf
  462}, 382 (2001).
\bibitem{MCEEP} P. Ulmer, MCEEP: Monte Carlo for Electro-Nuclear
  Coincidence Experiments, Version 3.5.
\bibitem{SIMC} J. Arrington, SIMC.
\bibitem{ryckebusch} S. Janssen, J. Ryckebusch, W. Van Nespen and
  D. Debruyne, Nucl. Phys. A {\bf 672}, 285 (2000).
\bibitem{bonn} R. Machleidt, K. Holinde and Ch. Elster,
  Phys. Rep. {\bf 149}, 1 (1987).
\bibitem{galster} S. Galster {\it et al.}, 
    Nucl. Phys. {\bf B32}, 221 (1971).
\bibitem{baselgmn} G. Kubon {\it et al.}, Phys. Lett. B {\bf 524}, 26
  (2002).
\bibitem{kopecky} S. Kopecky {\it et al.}, Phys. Rev. Lett. {\bf 74},
  2427 (1995).
\bibitem{lomon} E.L. Lomon, Phys. Rev. C {\bf 66}, 045501 (2002).
\bibitem{lightfront} F. Cardarelli and S. Simula, 
    Phys. Rev. C {\bf 62}, 065201 (2000).  
    Phys. Lett. B {\bf 467}, 1 (1999).
\bibitem{holzwarth} G. Holzwarth hep-ph/020138; Z. Phys. A {\bf 356},
  339 (1996) [Fit B2 is shown in Fig.~2].
\bibitem{pfsa} S. Boffi, L. Ya. Glozman, W. Klink, W. Plessas,
  M. Radici, and R.F. Wagenbrunn, Eur. Phys. J. A {\bf 14}, 17 (2002).
\bibitem{miller} G.A. Miller, Phys. Rev. C {\bf 66}, 032201(R) (2002).
\end{thebibliography}
\end{document}